# A Newer User Authentication, File encryption and Distributed Server Based Cloud Computing security architecture

Kawser Wazed Nafi[1,2], Tonny Shekha Kar[2], Sayed Anisul Hoque[3], Dr. M. M. A Hashem[4]

[1]Lecturer, Stamford University, Bangladesh
[2]Khulna University of Engineering and Technology
[3]Chittagong University of Engineering and Technology
[4]Professor, Khulna University of Engineering and Technology

*Abstract*— **The cloud computing platform gives people the opportunity for sharing resources, services and information among the people of the whole world. In private cloud system, information is shared among the persons who are in that cloud. For this, security or personal information hiding process hampers. In this paper we have proposed new security architecture for cloud computing platform. This ensures secure communication system and hiding information from others. AES based file encryption system and asynchronous key system for exchanging information or data is included in this model. This structure can be easily applied with main cloud computing features, e.g. PaaS, SaaS and IaaS. This model also includes onetime password system for user authentication process. Our work mainly deals with the security system of the whole cloud computing platform.**

## I. INTRODUCTION

At the present world of networking system, Cloud computing [1] is one the most important and developing concept for both the developers and the users. Persons who are interrelated with the networking environment, cloud computing is a preferable platform for them. Therefore in recent days providing security has become a major challenging issue in cloud computing.

In the cloud environment, resources are shared among all of the servers, users and individuals. As a result files or data stored in the cloud become open to all. Therefore, data or files of an individual can be handled by all other users of the cloud. [2, 3] Thus the data or files become more vulnerable to attack. As a result it is very easy for an intruder to access, misuse and destroy the original form of data. An intruder can also interrupt the communication. Besides, cloud service providers provide different types of applications which are of very critical nature. Hence, it is extremely essential for the cloud to be secure [4]. Another problem with the cloud system is that an individual may not have control over the place where the data needed to be stored. A cloud user has to use the resource allocation and scheduling, provided by the cloud service provider. Thus, it is also necessary to protect the data or files in the midst of unsecured processing. In order to solve this problem we need to apply security in cloud computing platforms. In our proposed security model we have tried to take into account the various security breaches as much as possible.

At present, in the area of cloud computing different security models and algorithms are applied. But, these models have failed to solve all most all the security threats. [5, 6, 7] Moreover for E-commerce [8] and different types of online business, we need to imply high capacity security models in cloud computing fields. Security models that are developed and currently used in the cloud computing environments are mainly used for providing security for a file and not for the communication system [9]. Moreover present security models are sometimes uses secured channel for communication [10]. But, this is not cost effective process. Again, it is rare to find a combined work of main server security, transaction between them and so on. Some models attempt on discussing about all of these, but are completely dependent on user approach. The models usually fail to use machine intelligence for generating key and newer proposed model. Some models have proposed about hardware encryption system for secured communication system [11]. The idea is usually straightforward, but the implementation is relatively difficult. Besides, hardware encryption is helpful only for the database system, not for other security issues. Authenticated user detection technique is currently very important thing. But, this technique is rarely discussed in the recently used models for ensuring security in cloud computing.

In this paper we have proposed new security architecture for cloud computing platform. In this model high ranked security algorithms are used for giving secured communication process. Here files are encrypted with AES algorithm in which keys are generated randomly by the system. In our proposed model distributive server concept is used, thus ensuring higher security. This model also helps to solve main security issues like malicious intruders, hacking, etc. in cloud computing platform. The RSA algorithm is used for secured communication between the users and the servers.

This paper is formatted in the following way: - section II describes related work of this paper work, section III describes proposed architecture and its working steps, section IV describes the experimental environment, results in different aspects and advantages of the proposed model, and section V describes the future aspects related to this paper work.

## II. RELATED WORK

Numerous research on security in cloud computing has already been proposed and done in recent times. Identification based cloud computing security model have been worked out by different researchers [12]. But only identifying the actual user does not all the time prevent data hacking or data intruding in the database of cloud environment. Yao's Garbled Circuit is used for secure data saving in cloud servers [13, 14]. It is also an identification based work. The flaw in this system is that it does not ensure security in whole cloud computing platform. Research related to ensuring security in whole cloud computing environments was already worked out in different structures and shaped. AES based file encryption system is used in some of these models [15, 16]. But these models keep both the encryption key and encrypted file in one database server. Only one successful malicious attack in the server may open the whole information files to the hacker, which is not desirable. Some other models and secured architectures are proposed for ensuring security in cloud computing environment [17, 18]. Although these models ensures secured communication between users and servers, but they do not encrypt the loaded information. For best security ensuring process, the uploaded information needs to be encrypted so that none can know about the information and its location. Recently some other secured models for cloud computing environment are also being researched [19, 20]. But, these models also fail to ensure all criteria of cloud computing security issues [21].

## III. PROPOSED MODEL

In our proposed model we have worked with the following security algorithms:-

- RSA algorithm for secured communication [22, 23]
- AES for Secured file encryption [24, 25, 26]
- MD5 hashing for cover the tables from user [27]
- One time password for authentication [28, 29].

At present ensuring security in cloud computing platform has become one of the most significant concerns for the researchers. We have undertaken these problems in our research, to provide some solution correlated with security. We have proposed the following security model for cloud computing data storage shown in Figure 1.

In this model, all the users irrespective of new or existing member, needs to pass through a secured channel which is connected to the main system computer. System server computer has relation with other data storage system. The data storage system can be servers or only storage devices. Here, each of the data storage devices can be thought as one or more servers in number. This means, there are no dedicated servers in cloud computing, rather all are independent servers and can be scaled as necessary.

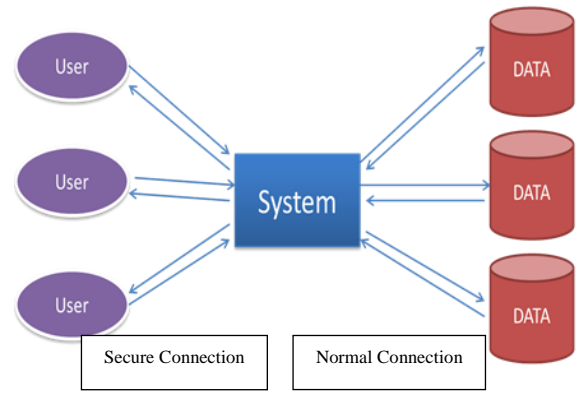

Figure 1.     Proposed Security Model

In the proposed model RSA encryption algorithm is used for making the communication safe. Usually the users' requests are encrypted while sending to the cloud service provider system. RSA algorithm using the system's public key is used for the encryption. Whenever the user requests for a file the system sends it by encrypting it via RSA encryption algorithm using the user's public key. Same process is also applied about the user password requests, while logging in the system later. After receiving an encrypted file from the system the user's browser will decrypt it with RSA algorithm using the user's private key. Similarly when the system receives an encrypted file from the user it will immediately decrypt it using its private key. As a result the communication becomes secured between the user and the system.

In the proposed security model one time password has been used for authenticating the user. The password is used to keep the user account secure and secret from the unauthorized user. But the user defined password can be compromised. To overcome this difficulty one time password is used in the proposed security model. Thus whenever a user login in the system, he/she will be provided with a new password for using it in the next login. This is usually provided by the system itself. This password will be generated randomly. Each time a new password is created for a user, the previous password for that user will be erased from the system. New password will be updated for that particular user. A single password will be used for login only once. The password will be sent to the users authorized mail account. Therefore at a same time a check to determine the validity of the user is also performed. As a result only authorized user with a valid mail account will be able to connect to the cloud system. By this system, existence of unauthorized user or a user with an invalid mail account will be pointed out. The newly generated password is restored in the system after md5 hashing. The main purpose of MD5 hashing is that this method is a one way system and unbreakable. Therefore it will be difficult for an unauthorized or unknown party for retrieving the password for a selected user even if gained access to the system database.

After connecting with the system a user can upload or download the file(s). For the first time when connected with the system the user can only upload file(s). After that users can both upload and download their files. When a file is uploaded by an user the system server encrypts the file using AES encryption algorithm. In the proposed security model 128 bit key is used for AES encryption. 192 bit or 256 bit can also be used for this purpose. Here the 128 bit key is generated randomly by the system server. A single key is used only once. That particular key is used for encrypting and decrypting a file of a user for that instance. This key is not further used in any instance later. The key is kept in the database table of the system server along with the user account name. Before inserting the user account name it is also hashed using md5 hashing. This insures that unauthorized person cannot retrieve the key to decrypt a particular file for a particular user by simply gaining access and observing the database table of the system server. As a result the key for a particular file becomes hidden and safe. Again when the encrypted file is uploaded for storing to the storage server, the path of the encrypted file along with the user account is kept and maintained in the database table on the storage server. Here user name is used for synchronization between the database tables of main system server and the storage server. The encrypted files on the storage server are inserted not serially. We have developed a hash table for determining where to insert a file into the database table. The algorithm for generating the hash table is described later in this section.

Login into the main system is compulsory when a user wants to download a previously stored file. When the user selects a file to download, the system automatically retrieves the key for the requested file from the main system server. The system matches user account name saved in its database table with that saved in the storage server after hashing it using md5 hashing. The path of the encrypted file from the storage server is found by using the user account name and the hash table input for the requested file.

In this model, the encryption key for a particular file of a particular user is only known to the main system server. The path of the encrypted file is only known to the storage server which is only known to the main server. For this, the key as well as the encrypted file is hidden from the unauthorized persons. In this communication system when a file is sent from the main system server to the storage server it is already in its fully encrypted form. That's why there is no need to provide security in this communication channel. At last, we propose hardware encryption for making the databases fully secured from the attackers and other unauthorized persons.

Figure 2 is the Pictorial representation of the proposed cloud security architecture. Here, single user and server represent n users and n servers.

An algorithm is developed, which is used for inserting the file in the main server (System), and in the database table where the encrypted file is kept. This is saturated from the system server for the cloud computing platform. In the system server, the file is inserted by maintaining the sequence. In file saving server, the file is inserted in a random order which becomes the output of the algorithm. The relations between the system server table and database server tables can be thought as disjoint sets. The pseudo code of the algorithm used is described in table I.

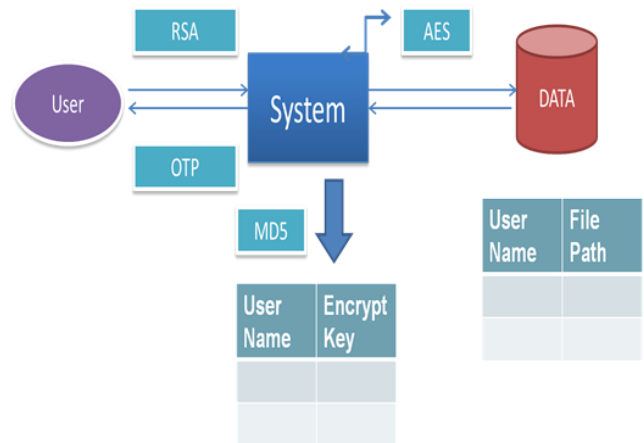

Figure 2. Proposed Security Model/ Structure

TABLE I New Algorithm for uploading file in the Proposed Cloud Architecture:-

The algorithm for generating the hash table which is used for inserting a file in the database table of the storage server is described below:

Step 1: - Select a seed $S$ for generating the hash table which is equal to the block size of the table. Block size means with how many positions of files will be taken from a series of execution

Step 2: - Compute the position where to insert a file.

Position = $N2 \bmod S$.

Where $N$ represents the no. of file and $S$ represents the seed value.

Step 3:- a) if Position is empty, then insert the file in that Position.

b) else, increment the Position and set Offset. Repeat step 3.

A sample hash table with seed S = 100 is shown in table II:

TABLE II Synchronization of files in two servers

| File No In System Server | Position Of File In Database Server | Offset |
|---|---|---|
| 1 | 1 | 0 |
| 2 | 4 | 0 |
| ⋮ | ⋮ | ⋮ |
| ⋮ | ⋮ | ⋮ |
| 5 | 25 | 0 |
| ⋮ | ⋮ | ⋮ |
| ⋮ | ⋮ | ⋮ |
| 15 | 26 | 1 |
| ⋮ | ⋮ | ⋮ |
| ⋮ | ⋮ | ⋮ |
|  |  |  |

## IV. EXPERIMENTAL RESULTS

In the lab we have worked with about 100 users and also with their files for studying and prove the efficiency of the proposed model. We have tried to find out different execution results which helped us to demonstrate our model with better result. Different conditions and positions were observed during the working and execution time of this proposed model.

### A. Lab Setup

- Platform: Visual Studio 2010 (asp.net)
- Processor: Core 2 Duo (2.93 GHz),
- RAM: 2 GB

In this environment, the whole model took average of 5 seconds for executing all the steps. This hardware configuration takes highest 2 seconds to encrypt about a 10 KB file. This model is fast enough and can be applied to current cloud computing environments.

### B. Case Studies

Working with the model in Lab at different times and with different user and their individual files, which are different from each other in size, contents, extension, etc. take different times for executing the overall model. Depending on the file size, program execution time varies from person to person. Among the 100 users result, 10 of them are shown in table III and table IV.

TABLE III    Execution time for Uploading File of 10 People

| Pers on No | File Size | Time Required for file Upload (Full Process) | Per son No | File Size | Time Required for file Upload (Full Process) |
|---|---|---|---|---|---|
| 1 | 1 KB | 3 sec | 6 | 17 KB | 10 sec |
| 2 | 4 KB | 5 sec | 7 | 15 KB | 10 sec |
| 3 | 14 KB | 9 sec | 8 | 5 KB | 5 sec |
| 4 | 7 KB | 6.5 sec | 9 | 2 KB | 3 sec |
| 5 | 9 KB | 8 | 10 | 8 KB | 8 sec |

TABLE IV    Execution time for Downloading File of 10 People

| Perso n No | File Size | Time Required for file Upload (Full Process) | Person No | File Size | Time Required for file Upload (Full Process) |
|---|---|---|---|---|---|
| 1 | 1 KB | 3.5 sec | 6 | 17 KB | 11 sec |
| 2 | 4 KB | 5.5 sec | 7 | 15 KB | 11 sec |
| 3 | 14 KB | 10 sec | 8 | 5 KB | 5.5 sec |
| 4 | 7 KB | 7 sec | 9 | 2 KB | 3.5 sec |
| 5 | 9 KB | 9 | 10 | 8 KB | 9 sec |

From table III and table IV we can see that the proposed model takes quite same time for execution like other present models. But it ensures higher security. Information is stored in main server about the databases where the encrypted files are kept. Thus, database encryption [30, 31] only in main server is enough so that no information is leaked. This makes the model cost effective and less time required for execution of the whole process. Secured information exchanging between the users and system gives protection of hiding information from the unauthorized users and intruders. Comparative analysis of the proposed model is shown in table V.

TABLE V    Advantages of the Proposed Model

| Points for discussion | Identific ation Based Model | File encryption based Model | Secured channel using model | Proposed Model |
|---|---|---|---|---|
| Ways of ensuring security | Only identify the authorized person, so hacker can get access on database | Key and file both remains in one server. So, getting access on one server helps to get all information | Intruder cant access the data, but uploaded file is not secured | Ensures security in data exchanging process. Only getting control over full system can leak information |
| **Points for discussion** | **Identific ation Based Model** | **File encryption based Model** | **Secured channel using model** | **Proposed Model** |
| Information leakage probability | Medium | Medium | Medium | Low |
| Complexity | Low | Medium | Low | Medium |
| Cost of establishing and maintaining | Low | Medium | High | Medium |
| Ensuring User Authenticatio n | Main theme | If key is chosen by user, then slightly authenticat e users | Probably not maintaine d | One time password system is used for user authenticatio n |
| Execution time | Small | Medium | Small | Medium |
| Security Breaking probability | Medium | Medium | Medium | Probably Low than others |

From the above comparative analysis, we can see that the proposed model works smoothly like others and ensures higher security than other present running models in a cloud computing environment.

## V. CONCLUSION

In this paper we have proposed a newer security structure for cloud computing environment which includes AES file encryption system, RSA system for secure communication, Onetime password to authenticate users and MD5 hashing for hiding information. This model ensures security for whole cloud computing structure.

Here, execution time is not subsequently high because implementation of each algorithm is done in different servers. In our proposed system, an intruder cannot easily get information and upload the files because he needs to take control over all the servers, which is quite difficult. The model, though it is developed in a cloud environment, individual servers' operation has got priority here. So, decision taking is easy for each server, like authenticate user, give access to a file etc.

In our proposed model we have used RSA encryption system which is deterministic. For this reason, it becomes fragile in long run process. But the other algorithms make the model highly secured. In future we want to work with ensuring secure communication system between users and system, user to user. We also want to work with encryption algorithms to find out more light and secure encryption system for secured file information preserving system.

## ACKNOWLEDGMENT

The Authors are willing to express their profound gratitude and heartiest thanks to all the researchers in the field of cloud computing architecture's security, specially to the developers of security algorithms, who have made their research work easy to accomplish.

## AUTHORS PROFILE

**Kawser Wazed Nafi** passed from Computer Science and Engineering department of Khulna University of Engineering and Technology in June 2012. He then started his career in Samsung Bangladesh R & D centre. Now he is working as Lecturer in Stamford University, Bangladesh in Computer Science and Engineering department. He has been working in Cloud Computing field for more than about one year. He has already published Journal and conference papers on Cloud Computing field in different knowable journals like IEEE, IJCA, IJCOT. His research interest is Cloud Computing, Ubiquitous Computing, Adhoc networks, Artificial Intelligence, Pattern Recognition and So on.

**Tonny Shekha Kar** passed from Computer Science and Engineering department of Khulna University of Engineering and Technology in June 2012. She has been working in Cloud Computing and Distributed Computing field for more than about one year. She has already published papers on Cloud Computing field, Patteren Recognition, etc in different knowable journals like IEEE, IJCA, IJCOT. Her research interest is Cloud Computing, ubiquitous Computing, Adhoc networks, Artificial Intelligence, Machine Learcning, Neural Network, Pattern Recognition and So on.

**Sayed Anisul Hoque** passed from Computer Science and Engineering department of Chittagong University of Engineering and Technology in March, 2011. He then started his career in Samsung Bangladesh R & D centre as software engineer. His research interest is Cloud Computing, Ubiqutous Computing, wireless network, operating system, android platform and So on.

**M. M. A. Hashem** received the Bachelor's degree in Electrical & Electronic Engineering from Khulna University of Engineering & Technology (KUET), Bangladesh in 1988. He acquired his Master's Degree in Computer Science from Asian Institute of Technology (AIT), Bangkok, Thailand in 1993 and PhD degree in Artificial Intelligence Systems from the Saga University, Japan in 1999. He is a Professor in the Department of Computer Science and Engineering, Khulna University of Technology (KUET), Bangladesh. His research interest includes Soft Computing, Intelligent Networking, Wireless

Networking, Distributed Evolutionary Computing etc. He has published more than 50 referred articles in international Journals/Conferences. He is a life fellow of IEB and a member of IEEE. He is a coauthor of a book titled "Evolutionary Computations: New Algorithms and their Applications to Evolutionary Robots," Series: Studies in Fuzziness and Soft Computing, Vol. 147, Springer-Verlag, Berlin/New York, ISBN: 3-540-20901-8, (2004). He has served as an *Organizing Chair, IEEE 2008 11th International Conference on Computer and Information Technology (ICCIT 2008) and Workshops, held during 24-27 December, 2008 at KUET.* Currently, he is working as a Technical Support Team Consultant for Bangladesh Research and Education Network (BdREN) in the Higher Education Quality Enhancement Project (HEQEP) of University Grants Commission (UGC) of Bangladesh.